\documentclass{rQUF2e}

\usepackage{subfigure}
\usepackage{bbm}
\usepackage{amsmath}
\usepackage{algorithm}
\usepackage[noend]{algpseudocode}
\usepackage{caption}
\usepackage{bm}
\usepackage{enumitem}

\usepackage{graphicx,epstopdf}

\makeatletter
\def\BState{\State\hskip-\ALG@thistlm}
\makeatother

\theoremstyle{plain}

\theoremstyle{definition}

\theoremstyle{property}
\newtheorem{property}{Property}

\theoremstyle{remark}

\usepackage{xcolor}

\usepackage{color,soul}

\begin{document}


\title{Portfolio Optimization with Sparse Multivariate Modelling}

\author{ PIER FRANCESCO PROCACCI $^1$ and TOMASO ASTE $^{1,2}$\\
\affil{$^1$Department of Computer Science, UCL, Gower Street, London, WC1E6BT, UK\\
$^2$Systemic Risk Centre, London School of Economics and Political Sciences, London,  WC2A 2AE, UK\
} \received{v1 released 28 Mar 2021} }

\maketitle

\begin{abstract}
Portfolio optimization approaches inevitably rely on multivariate modeling of markets and the economy. 
In this paper, we address three sources of error related to the modeling of these complex systems:
1. oversimplifying hypothesis; 
2. uncertainties resulting from parameters' sampling error; 
3. intrinsic non-stationarity of these systems. 
For what concerns point 1. we propose a $L_0$-norm sparse {elliptical} modeling and show that sparsification is effective.  
The effects of points 2. and 3. are quantified by studying the models' likelihood in- and out-of-sample for parameters estimated over train sets of different lengths. 
We show that models with larger off-sample likelihoods lead to better performing portfolios up to when two to three years of daily observations are included in the train set. For larger train sets, we found  that portfolio performances deteriorate and detaches from the models' likelihood, highlighting the role of non-stationarity. {We further investigate this phenomenon by studying the  out-of-sample  likelihood of individual observations  showing that the system changes significantly through time. Larger estimation windows lead to stable likelihood in the long run, but at the cost of lower likelihood in the short-term: the `optimal' fit in finance needs to be defined in terms of the holding period.}
Lastly, we show that sparse models outperform full-models in that they deliver higher out of sample likelihood, lower realized portfolio volatility and improved portfolios' stability, avoiding typical pitfalls of the Mean-Variance optimization.

\end{abstract}

\begin{keywords}
Portfolio Construction; Market States; Markowitz; Mean-Variance; Information Filtering Networks; TMFG; Sparse inverse covariance; Correlation Structure.
\end{keywords}

\begin{classcode}C38, G61, G15, G17 \end{classcode}

\section{Introduction}

Quantitative approaches to asset management have accumulated unprecedented popularity over the last few decades. 
Of all the algorithms and strategies developed, portfolio selection models are among those that have received wider attention. The essence of portfolio investing is to find the best way of assigning weights to a given portfolio of assets
to maximize future portfolio returns while minimizing the investment risk. 
 The exploration of this filed starts with Markowitz's Mean-Variance optimization process \citep{Markowitz52}. \\
The theory of mean–variance-based portfolio selection is still today a cornerstone of modern asset management. It rests on the presumption that rational investors choose among risky assets purely on the basis of expected return and risk, with risk measured as portfolio variance. The theoretical foundation of this framework is sound if either: investors exhibit quadratic utility, in which case they ignore non-normality in the data \citep{Gollier2001}, or all the higher moments of the portfolio distribution can be expressed as a function of mean and variance and hence all optimal solutions satisfy the mean-variance criterion.
Also, the``optimality" of the mean-variance portfolios is based on the assumption that investors live in a one-period world, while in reality they have an investment horizon that lasts longer than one period.
Markets indeed constantly change over time and investors are subject to inflows/outflows forcing them to adjust their allocation and take corrective actions.
\\
Form a general, high level, perspective all portfolio optimization approaches are based on a multivariate model of the variables in the market and the economy. The optimization strategies are devised to maximize profits and minimize risks based on such models. 
In modeling these complex systems there  are, however, several sources of inaccuracies and errors with the three main ones being:
1. oversimplifying hypothesis (such as the use of normal distributions); 
2. uncertainties resulting from the estimation of the parameters from datasets of limited sizes; 
3. intrinsic  non-stationarity of these systems, which makes in-sample estimations, based on past observations, inadequate for the estimation of off-sample, future properties.  
Most likely all three of these factors -- and others -- contribute to undermining the predictive power of any attempt of modeling markets.
\par
\bigskip
While a large deal of literature has been devoted to relaxing some of the most unrealistic model assumptions (point 1.) the current main pitfall of portfolio optimization is attributed to error maximization (point 2.). This effect has long been established in literature \citep{Michaud89, Nawrocki96}. Essentially, inputs into the mean-variance optimization are measured with uncertainty, and the optimization procedure tends to pick those assets which appear to have the most attractive features -- but these are outlying cases where estimation error is likely to be the highest, hence maximizing the impact of estimation error on portfolios' weights. 
The estimation error is also amplified by market evolution which makes the training on the past not fully representative of future market behavior (point 3.).
\par
\bigskip
In this paper, we address all three sources of inaccuracies.
For what concerns point 1. we propose a $L_0$-norm topologically regularized sparse {elliptical} modeling \citep{aste2020topological} and show that  sparsification is effective.  
We quantify the effects of estimation error and non-stationarity on portfolio performances (point 2. and 3.) by assessing the goodness of models' statistical likelihood for estimates over train sets of different lengths. 
Specifically, we study how the realized portfolio variance reacts to different out-of-sample likelihoods of the input parameters and, particularly, to sparse models. 
Further, we analyze how sparse precision matrices impact the magnitude and the stability of portfolio weights.
\par
\bigskip
The remainder of this paper is organized as follows: in Section \ref{lit_review} we briefly review the theory around portfolio construction, highlighting the pitfalls on assumptions and estimation error and the main solutions proposed in literature; in Section \ref{methodology} we outline our methodology and experiments design and in Section \ref{results_sec} we present the results. Appendices~\ref{elliptical_sec} and \ref{EllipticalProperties} are devoted to recalling some useful aspects of Elliptical distributions.

\section{Literature Review}
\label{lit_review}

\subsection{Modern Portfolio Theory}
\label{markowitz_sec}
Considering a portfolio of $n$ assets with weights $\bm{W}=(w_1,...,w_n)$, returns $\bm{R}=(\bm{r}_1,...,\bm{r}_n)$ and portfolio returns  
\begin{equation}
\bm{r}_p=\bm{WR}^T,
\end{equation}
the standard Mean-Variance optimization problem consists in minimizing the portfolios' variance $\sigma_{p}$ for fixed levels of expected returns $\mathbb{E}[\bm{r}_p]=\bar r_p$:

 \begin{equation}
\begin{aligned}
	\min_{\bm{W}} \quad & \sigma_{p}^{2}=\bm{W \Sigma W}^T \\
	\textrm{s.t.} \quad & \mathbb{E}[\bm{r}_p] = \bar r_p, \\
	\textrm{and} \quad & \bm{W}\mathbbm{1}=1, \\
  \label{min_variance}
\end{aligned}
\end{equation}
where 
$\bm{\Sigma}\in\mathbb{R}^{n\times n}$ is the assets' covariance matrix and $\mathbbm{1}\in\mathbb{R}^{n\times 1}$ is a basis column vector with all elements equal 1. 
Solving for $\bm{W}$ for different values of $\bar r_p$, one can obtain the optimal weights (i.e. the weights that minimize the portfolio variance) corresponding to different portfolio expected returns $\bar r_p$ yielding the so-called efficient frontier -- i.e. the set of optimal weights which provide the lowest variance for each level of expected return.
\par
\bigskip

As discussed in the introduction, this optimization is only concerned with the first two moments of the distribution of portfolios' returns and it does not deal with multiperiod investment decisions.
These pitfalls have largely been discussed in literature. \cite{Kritzman2000} provides a clear review of what are the assumptions under which repeatedly investing in one-period-efficient portfolios will also result in multiperiod efficiency. Also, many models have been proposed to deal explicitly with multiperiod optimality (see, for example, \cite{Ng00} or \cite{Arditti76}).
With respect to non-normality of returns, a rich literature is available on both alternative parametrization of data \citep{Bamberg01} and optimisation frameworks that consider other distribution moments \citep{Hogan74, Bawa78} or other measures of risk/return \citep{Harlow91, Lwin2017}.
\par
\bigskip

From  the optimization problem in Eq.\ref{min_variance} it is also clear that the optimization does not treat the error and uncertainty around the parameters $\bm{\Sigma}$ and $\bm{\mu}$. The difference between the estimated and true distribution parameters is called estimation error. It arises from both the sampling procedure or availability of data and non-stationarity. The error coming from sampling, also referred to as sampling error, is due to parameters used in the portfolio optimization process being typically point estimates -- we can only expect these estimates to equal the true distribution parameters if our sample is infinitely large. {Assuming stationary data, sampling error could be fixed by increasing the number of observations in the estimation sample. Indeed, the convergence rate is in the inverse of the square-root of the sample size, as dictated by the law of large numbers. This would come handy in our times of increasing data availability.} However, a second source of estimation error comes from non-stationarity. A time series is said to be non-stationary if its distribution parameters (or the distribution itself) changes over time -- in this case, extending the length of observations might reduce the contribution of sampling error to estimation error, but at the same time, it could increase that of non-stationarity \citep{Broadie1993}.

\par
\bigskip
Many techniques have been proposed in literature to deal with this phenomenon, both relying on heuristic methods and decision-theoretic foundations \citep{SchererRisk}. Heuristic approaches mainly propose to constrain the optimization problem in order to impose feasible optimal weights.
\par 
\cite{Resampling} address explicitly the sampling error proposing a Monte Carlo based procedure called resampling. In order to model the randomness of the input mean vector and covariance matrix, portfolio resampling consists in repeatedly drawing from the return distribution given by the point estimates and creating $n$ artificial new samples. For each sample, an efficient frontier is estimated and the final, resampled-efficient frontier is given by the average weight across all of the resampled portfolios. 
\par
\par
From a decision-theoretic  perspective, Bayesian techniques have recently played a primary role in literature. The rationale behind Bayesian statistics for portfolio construction is to include non-sample information to tackle the effect of parameter uncertainty on optimal portfolio choice. Instead of a point estimate, Bayesian approaches produce a density function for the parameters involved, by combining sample information (likelihood) with prior belief, potentially coming from non-sample information. A special case of this general approach is the seminal work of \cite{Black92}. In their pioneering work, the authors assume assets' returns to be normally distributed with mean equal to the ``equilibrium returns" (that is, the mean returns that would return the market portfolio if used in a mean-variance optimization) and combine this ``sample" information with investors' views on the assets. In this way, in absence of an informative prior from investors, the model would return the market or ``equilibrium" portfolio. In presence of investors' priors, instead, the allocation would diverge from the equilibrium portfolio accounting for investors' views and proportionally to their confidence level.
Other than being highly appealing from a practitioner's perspective, the model proposed in \cite{Black92} highlights the flexibility of the Bayesian framework, with many sources of information that could potentially be used in combination or to update the in-sample information. This is a very active area of research with recant notable examples including \citep{BayesianMethodsInInvesting} and \citep{Defranco19}.

\subsection{Precision matrix estimation with topological regularization}
The literature concerning parameter estimation has greatly developed in recent years. While the increasing availability of data has represented the fuel for data-greedy machine learning models, it has also exacerbated problems related to the curse of dimensionality and overfitting.
In essence, the goal is to extract the largest amount of information from data with a model that avoids overfitting, generalizing well with new data. {Furthermore, model interpretability (or explainability) has become increasingly important and for this purpose, simpler and sparser models with a smaller number of parameters are preferred.} To this extent, understanding the dependency structure among variables has proven to be essential to capture the collective behaviour of the systems and many techniques have been introduced in literature. Examples are dimensionality reduction methods \citep{Maaten2009} or shrinkage techniques \citep{Ledoit20, Friedman_GLasso}.
\par
\bigskip

One possible approach to represent the set interactions in a complex system is to model them as a network structure where the vertices are the system's elements and edges between vertices indicate the interactions between the corresponding elements \citep{lauritzen1996}. Information filtering networks \citep{Aste_Tool} aim at retrieving the relevant sub network of interactions among the elements of the system. In the pioneering work of \cite{Mantegna1999}, it was proposed to investigate financial systems by the extraction of a minimal set of relevant interactions associated with the strongest correlations belonging to the Minimum Spanning Tree (MST). The MST structure is, however, a drastic filtering tool and is likely to discard valuable information. \cite{Aste_Tool} and \cite{Aste_Dynamical} expanded the concept by introducing the general idea of information filtering networks. In particular, they show that graphs of different complexities can be constructed by iteratively linking the most strongly connected nodes under the constraint of generating \textit{planar} or \textit{hyperbolic} graphs. There is now a large body of literature proving network filtering to be a powerful tool to associate a sparse network to a high-dimensional dependency measure with applications ranging from financial markets \citep{Aste_Parsimonious} to biological systems \citep{Aste_biology} and econophysics \citep{Mantegna2000}. 
Recent developments extend the information filtering network methodology to chordal graphs \citep{Aste_Parsimonious,AsteTMFG,massara2019learning}. This allows to directly associate the information filtering network with a positive definite sparse inverse covariance matrix that is a $L_0$-norm topological regularization of the full covariance estimate \citep{aste2020topological}. 
\par
\bigskip

In our analysis, we use the TMFG-LoGo network filtering approach \citep{AsteTMFG,Aste_Parsimonious}. TMFG-LoGo approach has proven to be more efficient and better performing, particularly when few data are available \citep{Aste_Parsimonious,aste2017causality}, with respect to  $L_1$-norm sparsification techniques such as GLASSO \citep{Friedman_GLasso}. Moreover, the TMFG procedure is computationally highly efficient and this is well suited in our resampling experiment.

\par
\bigskip
Despite the rich literature on portfolio construction techniques, to the best of our knowledge, the link between a measure of estimation goodness and portfolio performances is still not specifically treated in the literature.

\section{Methodology}
\label{methodology}
The estimation error is quantified by measuring how well the functional form of the multivariate probability distribution, $f_{\bm{\theta}}(\bm{X})$ defined via the parameters $\bm{\theta}$ estimated in-sample, describes the actual data out-of-sample. The statistical measure that describes how likely are  observations to belong to the estimated probability function is the likelihood. The likelihood principle is a cornerstone of statistical analysis in that maximum likelihood estimators are guaranteed to be asymptotically efficient under mild conditions (\cite{wald49}, \cite{Cramer46}, \cite{Daniels61}). 
In this section, we introduce our methodology and discuss our results for the multivariate normal case. 
{In Appendix \ref{elliptical_sec} we discuss  further the generality of this approach and show that it extends to other distributions of the elliptical family including, in particular, the multivariate Student-t.}
\par
\bigskip

The logarithm of the likelihood for the normal case is proportional to:
\begin{equation}
	\ln \mathcal{L}(\bm{\theta; X}_t) = \ln \left| \bm{J} \right| - (\bm{X}_t- \bm{\mu})\bm{J}(\bm{X}_t- \bm{\mu})^T + k \;\;,
  \label{likelihood_firstDef}
\end{equation}
where  $\bm{X}_t=(x_{t,1},x_{t,2},...,x_{t,n})$  is the $n$-dimensional multivariate returns observation vector at time $t$; $\bm{\theta}$ is the model parameter set, which includes $\bm{\mu}$ the vector of means and $\bm{J}$  the generalized precision matrix and; $k$ is a constant which is independent from ${\bm{\mu, J}}$ or $\bm X_t$ (see Appendix \ref{elliptical_sec}).
In the multivariate normal case  $\bm J = \bm \Sigma^{-1}$ is the inverse of the covariance. 
{Our results generalize to other elliptical distributions with defined covariance where $\bm J$ is proportional to the inverse covariance, which we assume is defined and invertible. }
Results for the Student-t are explicitly reported in appendix \ref{elliptical_sec}. 

\bigskip
Our goal is to study the log-likelihood in Eq. \ref{likelihood_firstDef} using different estimation windows and comparing how the precision matrices, estimated through maximum-likelihood and TMFG-LoGo, perform.
We considered a {\bf dataset} of daily closing prices of US stocks entering among the constituents of the S\&P 500 index between 02/01/1997 and 31/12/2015. After screening for those continuously traded and those not displaying abnormal returns, we reached a final dataset of 342 stocks. For each asset $i = 1, ..., n$, we calculated the corresponding daily returns  $x_{t,i} = P_{t,i}/P_{(t-1),i}-1$, where $P_{t,i}$ is the closing price of stock $i$ at time $t$, for a total of 4026 daily multivariate observations.
\par
\begin{figure}
\begin{center}
\begin{minipage}{140mm}
\begin{center}
\includegraphics[scale=0.50]{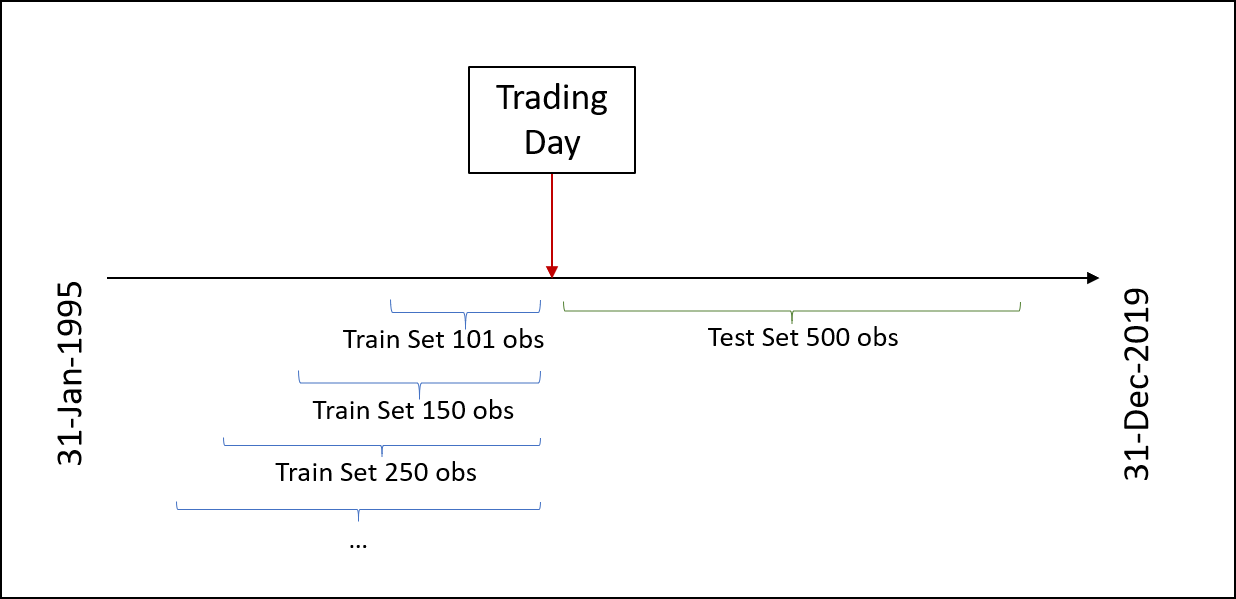}
\caption{Training and Testing scheme. We randomly sample the ending date of the training period, the `trading day'. We then estimate the model parameters considering different training windows using observations up to the randomly selected trading day. The following 	500 observations are used for testing.
\label{Training_Sketch}}
\end{center}
\end{minipage}
\end{center}
\end{figure}

\bigskip
We designed a resampling experiment in which we  select 100 stocks at random among the 342 and a random trading day spanned in our dataset. Starting from the randomly selected trading day and going back in time, we define five train sets of different sizes by including an increasing number of observations. We start at 101 observations, then 150, 250, 500, 1000 and finally 1500 observations. We then use a fixed-length test set of 500 observations following the randomly selected trading day. We keep the test set length fixed to avoid biases and selected 500 observations so that, for all estimation windows, the main crisis event (i.e. Global Financial Crisis in 2008) can be randomly included in or out of sample. Figure \ref{Training_Sketch} shows a sketched example of our train/test split with different estimation windows.
We use the train set to estimate the mean vector $\bm{\mu}$, the maximum-likelihood covariance matrix $\bm{\Sigma}$ and the sparse TMFG LoGo  covariance matrix $\bm{\Sigma_{TMFG}}$. These parameters are then used to compute the log-likelihood in eq. \ref{likelihood_firstDef}  for both  in-sample and out-of-sample observations. 
We then investigate how the different estimates used in a portfolio optimization procedure affect the optimal weights and portfolio characteristics.
To this extent, we considered the standard, unconstrained Markowitz optimization problem described in Section~\ref{markowitz_sec}. This is done to avoid any bias coming from constraints in our analysis and to keep the framework as plain as possible. We focus therefore our analysis on the minimum variance portfolio, that is the efficient portfolio that minimizes the expected variance. To obtain the solution for the minimum variance portfolio, the portfolio optimization problem in eq.~\ref{min_variance} rewrites:
\begin{equation}
\begin{aligned}
	\min_{\bm{W}} \quad & \sigma_{p}^{2}=\bm{W \Sigma W}^T \\
	\textrm{s.t.} \quad & \bm{W}\mathbbm{1}=1, \\
\end{aligned}
\end{equation}
which gives the optimal, minimum variance weights:
\begin{equation}
	\bm{W_{min}^*} = c \; \mathbbm{1}\bm{\Sigma^{-1}} ,
  \label{min_var_port}
\end{equation}

where $c=\frac{1}{\mathbbm{1}^T \bm{\Sigma^{-1}}\mathbbm{1}}$ is a normalization constant.
Considering the estimation scheme described above and outlined in fig.~\ref{Training_Sketch}, the estimated covariance matrices are used as inputs in the minimum variance optimal portfolio, eq. \ref{min_var_port}, to compare the different log-likelihood levels obtained out-of-sample and the corresponding effects on portfolio performances.

\section{Results}
\label{results_sec}

\subsection{Likelihood Comparison}
\label{lik_res}
\begin{figure}[h]
\begin{center}
\begin{minipage}{140mm}
\begin{center}
\subfigure[Likelihood comparison in-sample]{
\resizebox*{6cm}{!}{\includegraphics{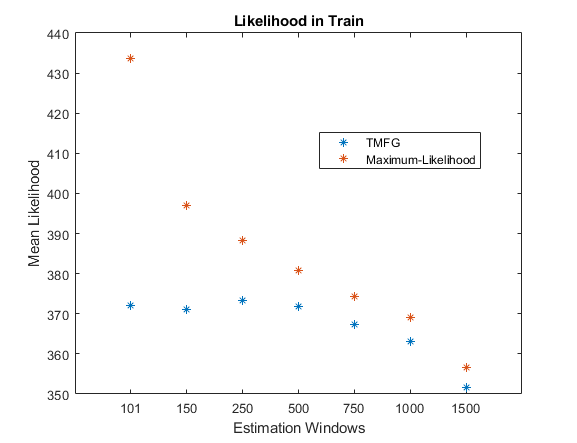}}\label{Likelihood_InTrain}}
\hskip.5cm
\subfigure[Likelihood comparison out-of-sample. Note the y axis break to fit the scale for 101 days estimation window.]{
\resizebox*{6cm}{!}{\includegraphics{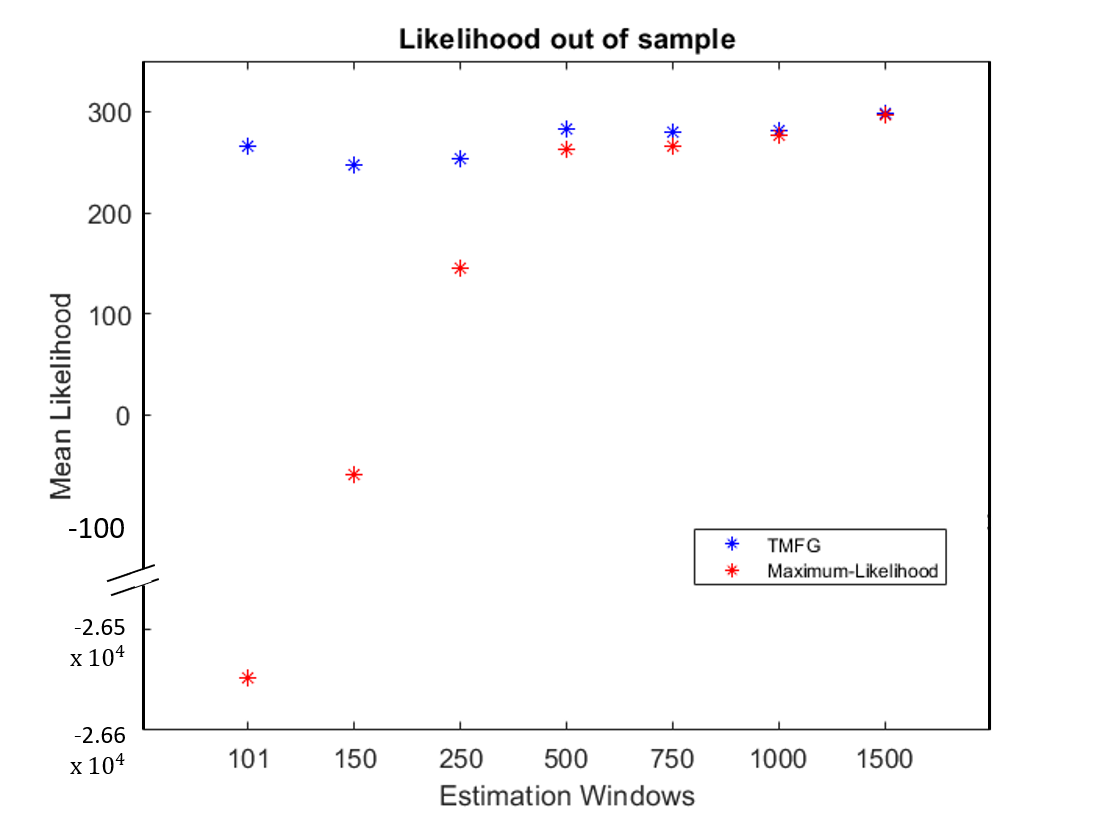}}\label{Likelihood_InTest}}
\caption{Log Likelihood computed in sample and out of sample. Both the likelihood computed using the maximum likelihood and the TMFG covariances decrease in sample as the sample size increases. The maximum likelihood covariance delivers by construction the highest likelihood, but the TMFG likelihood tracks it closely. In test, instead, the TMFG covariance always attain the highest likelihood and delivered good results also when the number of observations becomes close to the number of variables.
\label{Likelihood_comparison}}
\end{center}
\end{minipage}
\end{center}
\end{figure}

Figure \ref{Likelihood_comparison} reports the average log-likelihood for the train data (fig. \ref{Likelihood_InTrain}) and for the test data (fig. \ref{Likelihood_InTest}) computed  across 100 resamplings. The larger the log-likelihood is, the better the parameters $\bm{\theta}$ are at describing the data, for the assumed model. Fig. \ref{Likelihood_InTrain} shows that, as expected and by definition, the maximum-likelihood estimate of the covariance matrix provides a higher in-sample likelihood as compared to the TMFG covariance, although the latter tracks quite closely the maximum-likelihood. Also, one might observe that the likelihood is strictly decreasing with the number of observations included in the estimation window. Indeed, as the number of observations decreases relative to the parameters, the model overfits the sample yielding larger in-sample likelihoods. Filtering the covariance matrix and reducing the number of parameters clearly limits the overfitting potential of the model as shown by the lower levels of likelihood attained by TMFG when fewer observations are used which, therefore, results in a larger gap in likelihood relative to the maximum-likelihood covariance.
\par
Perhaps more interestingly, fig. \ref{Likelihood_InTest} reports the likelihoods obtained out-of-sample using the two different in-sample estimates of the covariance matrix. The first observation is that TMFG-LoGo provides a substantially larger log-likelihood, especially for short estimation windows. This result is exacerbated by the fact that when 101 observations are considered, {the number of stocks is very close to the number of observations in our samples. While the resulting covariance is still full-rank ({\it number of observations $>$ number of variables}), it leads to unstable estimation in the maximum-likelihood covariance (i.e. the so-called ``the curse of dimensionality'') whereas TMFG-LoGo is still well defined.} Note that there is a break y axis of the figure to allow a better inspection of the results. The figure shows that for longer estimation windows, the out-of-sample log-likelihood computed with the maximum-likelihood covariance tends to converge to the TMFG likelihood which, however, a) always provides the best out-of-sample likelihood in our experiment and b) provides quite stable likelihood values also for shorter estimation windows.  We conclude that the TMFG-LoGo algorithm does a good job at filtering the correlation structure providing higher out-of-sample likelihood and stable results with shorter estimation windows, confirming the results with stationary time series previously reported in \citep{Aste_Parsimonious}

\subsection{Impact of precision matrix estimate on optimal portfolios}
\begin{figure}[H]
\begin{center}
\begin{minipage}{140mm}
\begin{center}
\includegraphics[scale=0.50]{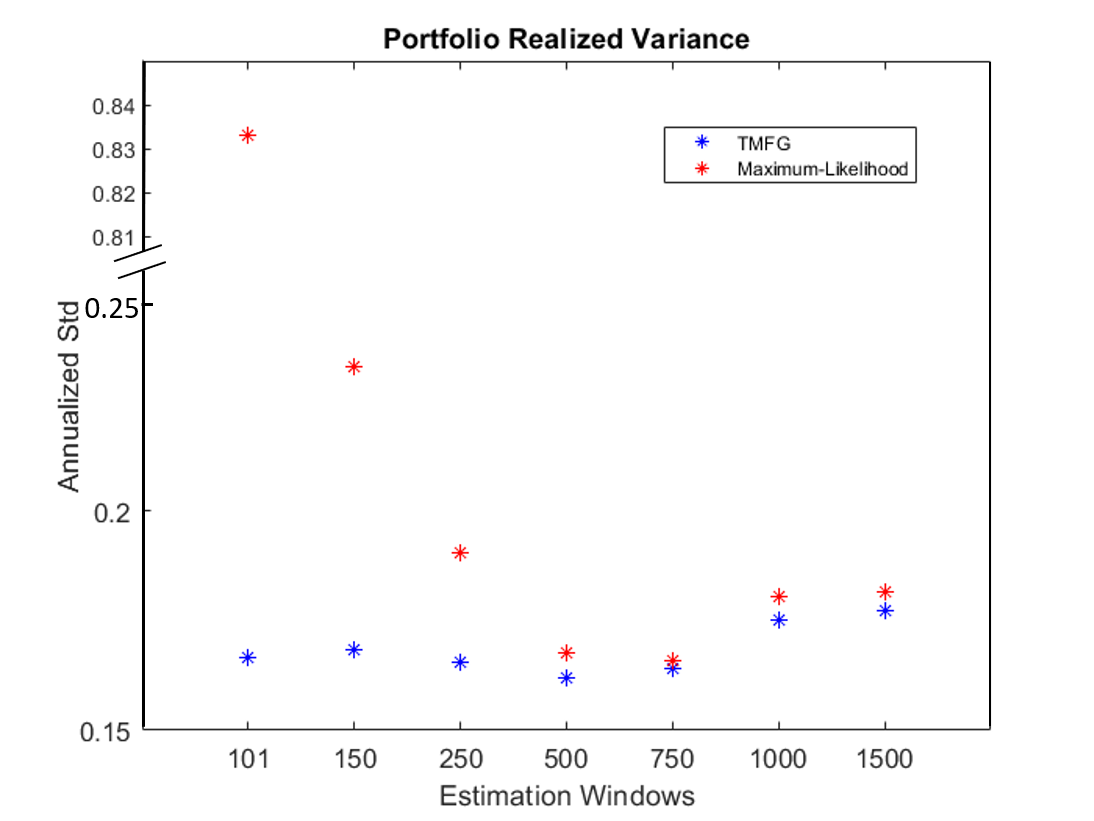}
\caption{Realized Standard Deviation. Increasing the estimation window and for higher values of Likelihood (figure \ref{Likelihood_comparison}), the realized standard deviation of portfolios decreases. Y axis break to fit the scale for 101 days estimation window..
\label{Port_Variance}}
\end{center}
\end{minipage}
\end{center}
\end{figure}

We now address empirically the question of what is the impact of different parameter estimates on portfolios weights and performances, when these parameters are used as inputs in the portfolio optimization problem in eq. \ref{min_variance}. Having focused our attention on the minimum variance portfolio on the efficient frontier, we report in Figure \ref{Port_Variance} the realized standard deviation of portfolios obtained using the same parameters which provided the log-likelihood displayed in Figure \ref{Likelihood_comparison}. The chart shows that, overall, the out-of-sample portfolio variance decreases as the likelihood increases up until when 750 observations are used. This is coherent with respect to the likelihood results that reported, indeed, increasing likelihoods for the same estimation windows. In particular, for shorter estimation windows, the TMFG-LoGo covariance matrix provides portfolios with significantly lower realized variance. Also, little changes are observed in the realized variance when observations from 101 to 750 are included, signalling that the TMFG-LoGo extract the relevant dependency links also when few observations are available.
The gap in performance tends to reduce as the number of observations in the estimation window increases, with the TMFG-LoGo portfolios always displaying lower volatility. However, when more than 750 daily observations are included, while the out of sample likelihood remains flat or slightly increases, the portfolios' variance tends to increase. 

\begin{figure}[h]
\begin{center}
\begin{minipage}{150mm}
\begin{center}
\subfigure[Realized volatilities obtained with Maximum Likelihood covariances]{
\resizebox*{6cm}{!}{\includegraphics{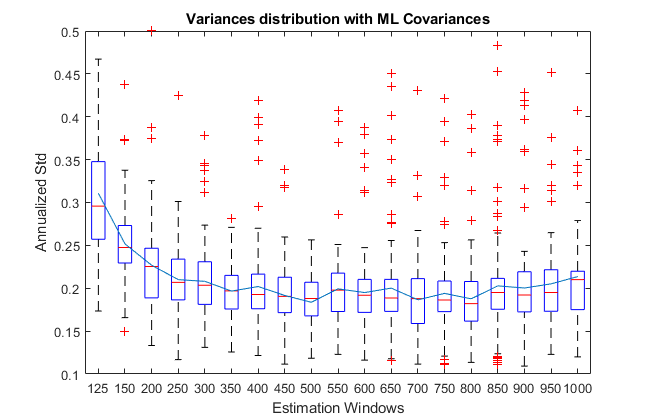}}\label{Long_Positions}}
\hskip.5cm
\subfigure[Realized volatilities obtained with TMFG filtered covariances]{
\resizebox*{6cm}{!}{\includegraphics{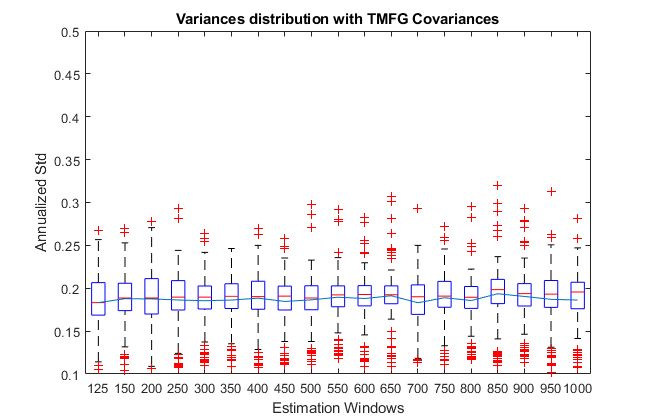}}\label{Short_Positions}}
\caption{Portfolio realized volatility across resamplings for different estimation windows. The box-plot shows the distribution of the variances obtained for 100 resampled portfolios and the the blue line overlaid shows the average variance (i.e. the mean of the variances' distribution).
\label{Variance_box}}
\end{center}
\end{minipage}
\end{center}
\end{figure}

{To further investigate this pattern, we report in Figure~\ref{Variance_box} the volatilities for all 100 resamplings and considering steps of 25 observations in the estimation windows. The figure confirms that the TMFG-LoGo covariances delivered overall less volatile portfolios across resamplings and estimation windows. Secondly, the figure shows that the portfolios obtain the lowest out-of-sample variance when approximately 2 to 3 years of daily observations (450 to 700 observations) are included in the train set. This pattern is clear for the Maximum likelihood portfolios, with means, quintiles and outliers drifting upwards when more than 750 observations are included. The TMFG filtered covariance regularizes and smooths this effect as well, but still when more than 750 observations are included, the resulting portfolios exhibit a slightly higher variance.
This is consistent with the literature showing that longer estimation windows provide worse forecasts in financial time series due to the regime-changing nature of financial markets \citep{Procacci2019}.
It is also worth emphasizing that the same mean vectors are used as inputs in the Markowitz optimization for both the TMFG-LoGo and the Maximum-Likelihood portfolios, hence the differences in performances are due solely to the different estimates of the covariance matrix.
} \\

\begin{figure}[h]
\begin{center}
\begin{minipage}{140mm}
\begin{center}
\subfigure[Number of ``Buy" positions.]{
\resizebox*{6cm}{!}{\includegraphics{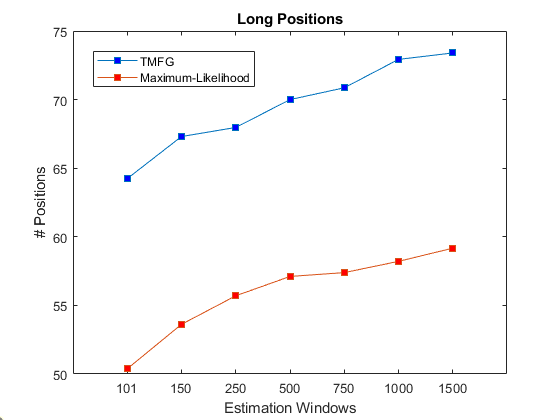}}\label{Long_Positions}}
\hskip.5cm
\subfigure[Number of ``Sell" positions.]{
\resizebox*{6cm}{!}{\includegraphics{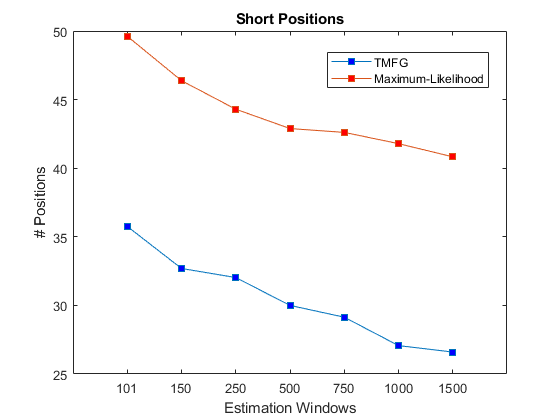}}\label{Short_Positions}}
\caption{Comparison of Buy/Sell Active Positions. As the number of training observations increases, the optimizations delivers an increasing number ``Long" positions. This tendency is anticipated when using TMFG fiiltered covarinace which always delivers an higher number of Long positions.
\label{Active_Positions}}
\end{center}
\end{minipage}
\end{center}
\end{figure}

\begin{figure}[H]
\begin{center}
\begin{minipage}{140mm}
\begin{center}
\subfigure[Distribution of optimal weights using Maximum-Likelihood covariance]{
\resizebox*{6cm}{!}{\includegraphics{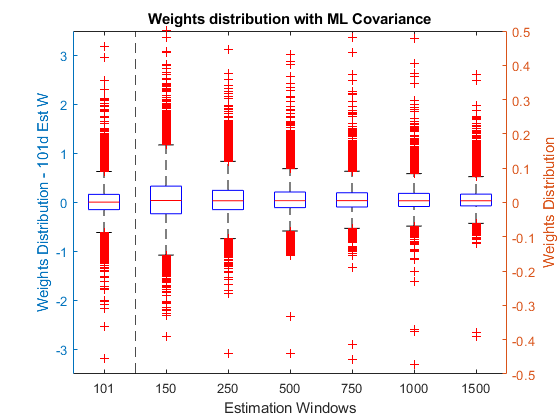}}\label{Long_Positions}}
\hskip.5cm
\subfigure[Distribution of optimal weights using TMFG covariance]{
\resizebox*{6cm}{!}{\includegraphics{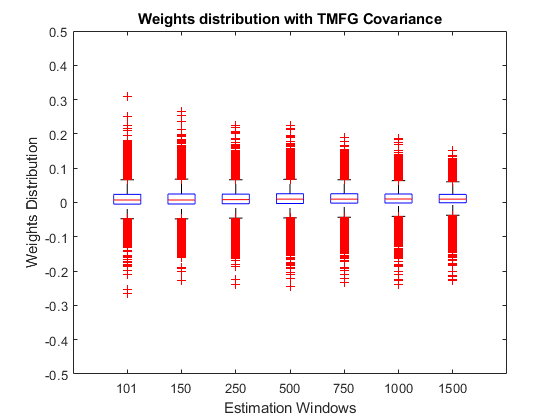}}\label{Short_Positions}}
\caption{Optimal Weights Distribution. Using the TMFG filtered covariance in the optimization provides stable weights as compared to the maximum-likelihood covariance, avoiding ``corner solutions" and enhancing diversification. 
\label{Weights_dist}}
\end{center}
\end{minipage}
\end{center}
\end{figure}

Finally, we address the impact of sparsity on portfolio construction by analyzing what is the effect of a sparse precision matrix on the active bets obtained from an optimization procedure. Figure \ref{Active_Positions} reports the number of Long (fig. \ref{Long_Positions}) and Short (fig. \ref{Short_Positions}) positions (i.e. positive and negative weights assigned to the stocks in portfolio) on average across the 100 resamplings. The first observation is that the number of long positions tends to increase as the estimation window increases and coherently the short positions diminish accordingly. Using TMFG-LoGo precision matrices anticipates this behavior, in that these portfolios always display a greater number of long positions also for short estimation windows. {The intuition behind this phenomenon is that over the long term, markets tend to post positive returns and possible outliers in assets' means and correlations are polished.} This  intuition is confirmed  by looking at the distribution of weights across resamplings in Figure \ref{Weights_dist}. This chart (note the different scales) shows that using the TMFG-LoGo covariance matrix significantly improves the stability of the optimal solutions, reducing outliers and avoiding ``corner", i.e. extreme solutions which are a typical pitfall of the unconstrained Markowitz optimization.
This results shows that the correlation coefficient among assets plays an important role in that for high correlation levels, the optimization procedure would prefer one stock in place of another for slightly more appealing mean or variance features. Having filtered the correlation structure in the TMFG-LoGo procedure, we obtained a portfolio that is much more general (hence the anticipated larger number of Long positions) and less sensitive to single assets features given the filtered correlation among stocks.
\par

Lastly,  considering the standard unconstrained optimization problem in eq.~\ref{min_variance}, both the maximum-likelihood and the TMFG matrices produce portfolios that are in the vast majority of cases investing in all assets. In other words, even considering a sparse precision matrix like in the TMFG-LoGo case, we very rarely found weights equal to zero assigned to some assets.

\subsection{Non Stationarity}

\begin{figure}[H]
\begin{center}
\begin{minipage}{150mm}
\begin{center}
\subfigure[Mean Likelihood for each observation across resamplings. Comparison of likelihoods obtained when 125, 750 and 1500 days are used in the train set.]{
\resizebox*{6.5cm}{!}{\includegraphics{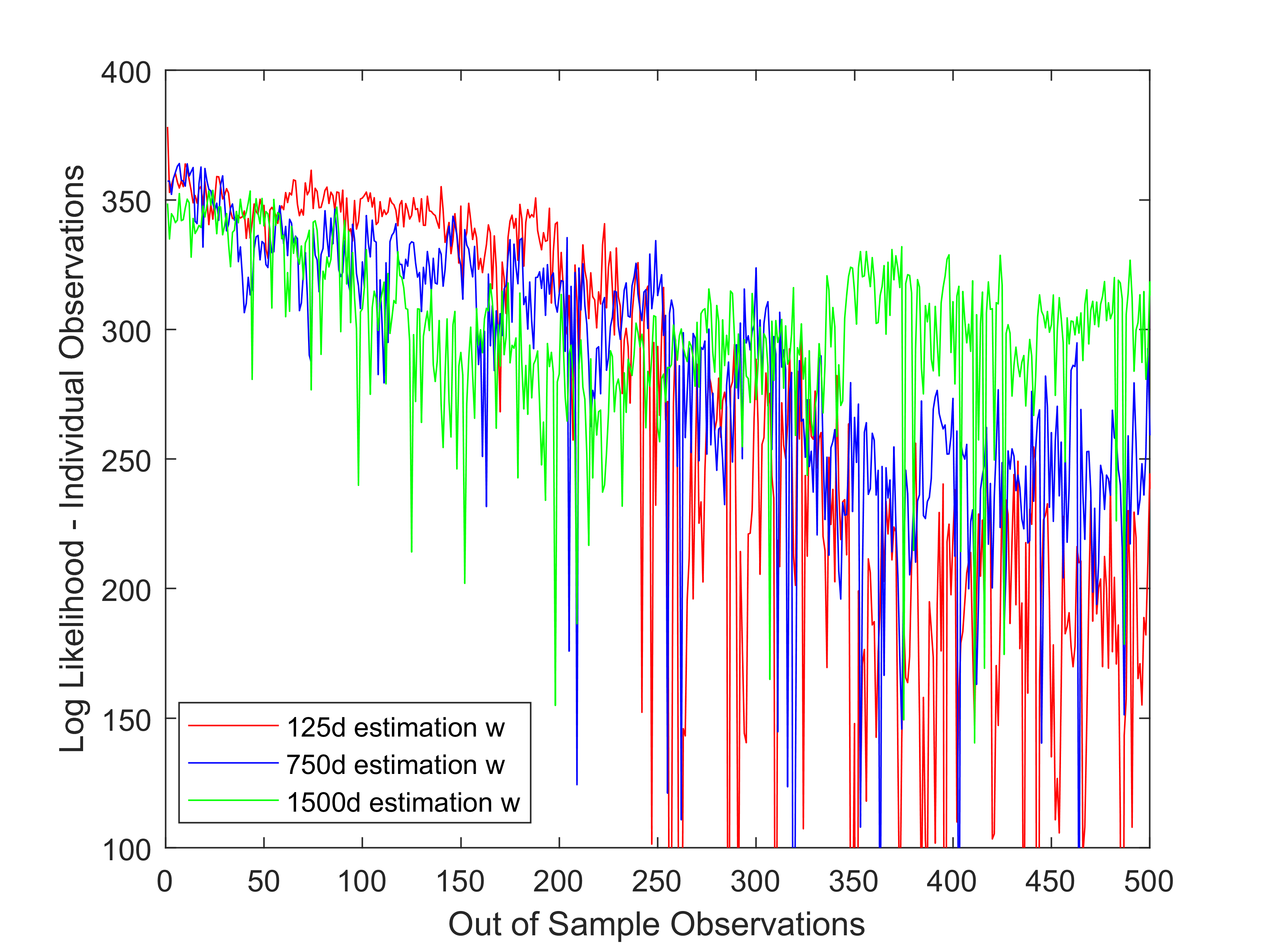}}\label{likObs}}
\hskip.6cm
\subfigure[Boxplot of likelihoods representing the quartiles and min-max levels for each observation across resamplings, having removed outliers. The plot is for 750 days train set (blue plot on the left).]{
\resizebox*{6.5cm}{!}{\includegraphics{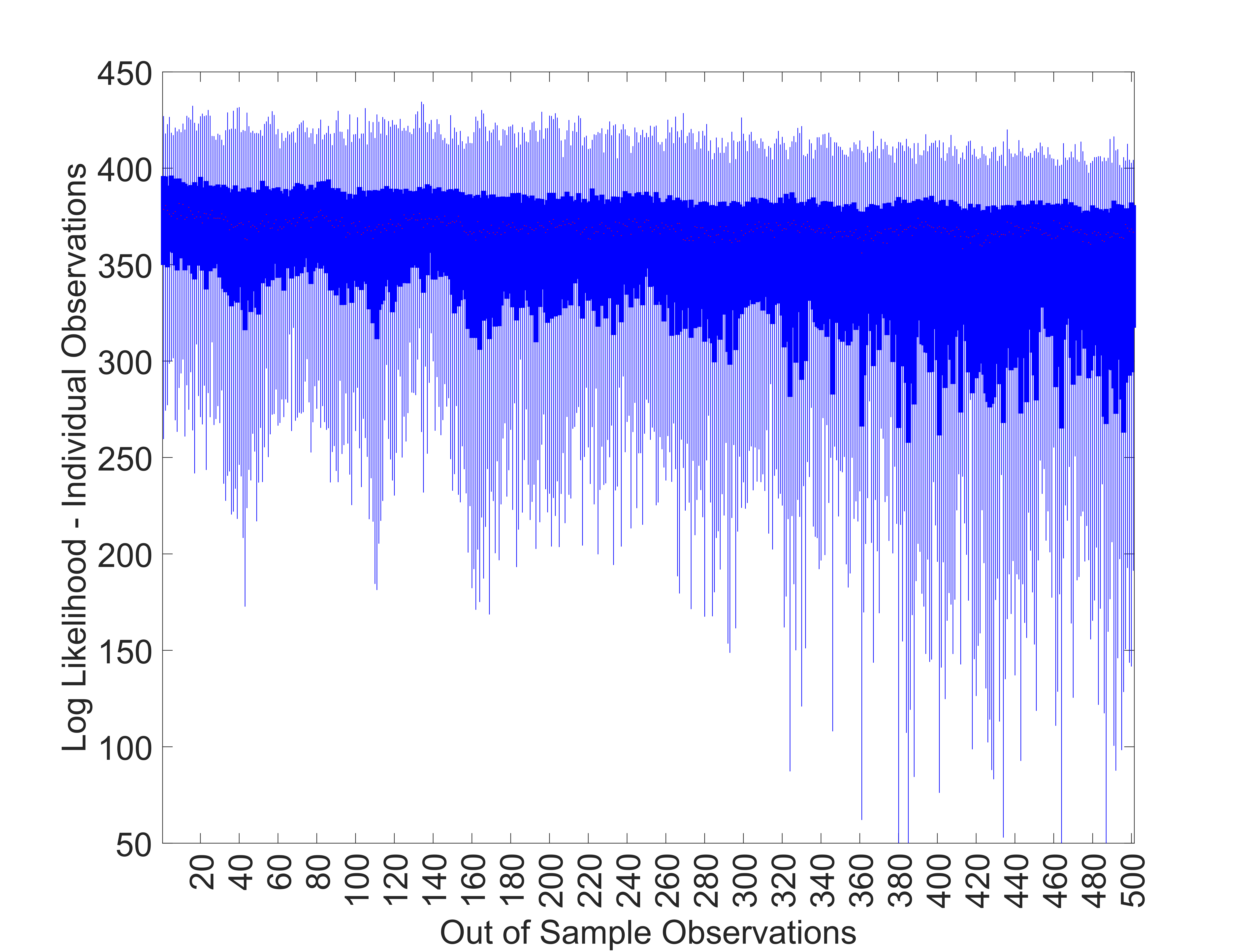}}\label{boxplotLik}}
\caption{Out-of-sample likelihood measured observation-by-observation.
\label{obsLikelihood}}
\end{center}
\end{minipage}
\end{center}
\end{figure}

From the results discussed in previous section and shown in Figure~\ref{Port_Variance} and Figure~\ref{Likelihood_comparison}, we found that the portfolio performances improve coherently with the likelihood up until approximately 3 years of observations are used in the train set to estimate the models parameters. However, when more onservations are included in the train set, the likelihood of the parameters detaches from the portfolio performances and we speculate that this is due to the role of nonstationarity. To further investigate this phenomenon, Figure~\ref{obsLikelihood} reports the likelihood corresponding to each out-of-sample observation in our experiment.

Figure~\ref{likObs} shows the average likelihood across 100 resamplings for each out-of-sample observation. We note that when shorter estimation windows are used to estimate the models' parameters (i.e. 125 days)  the likelihood is higher in the days immediately following the estimation window, but tends to rapidly decrease as the observations depart from the training window. Larger estimation windows (i.e. 750 or 1,500 days)  instead, lead to a more stable likelihood in the long run, but at the cost of a lower likelihood for the obervations closer to the estimation set. Figure~\ref{boxplotLik} shows the observation-wise box-plot of the likelihood computed across the resamplings when 750 observation are used in the train set. The box plot reports the 25\%-75\% quantile interval (dark blue) and the max-min interval (`whiskers' light blue) having excluded the `outliers' that are below the whiskers' \citep{Langford06}. Other than decreasing means, the figure shows that as the observations depart from the train set, the amount of observations posting a significantly lower likelihood increases, together with the downside volatility. 
In other words, it is more likely to have observations that are far from the model estimated in sample, supporting the conclusions drawn from Figure~\ref{likObs}.
\par
\bigskip
As we discussed in \citep{Procacci2019}, market states tend to be persistent in daily observations. Shorter estimation windows, therefore, tend to better describe the system belonging to the same `state' which is likely to be persistent for adjacent observations.
Notice that by considering the aggregate behaviors across 100 resamplings, we want to avoid specific market conditions and state shifts, but rather focus on the general behaviour. The evolution of the financial system is obviously very dynamic and the goodness of parameters is certainly dependent on both systemic and idiosyncratic events.
\par
These results also provide further insights on the findings discussed in Figure~\ref{Likelihood_comparison} and Figure~\ref{Port_Variance} in that our conclusions are dependent on the number of out-of-sample observations that in our case coincides with the portfolio holding period - i.e. 500 days in our experiments. Shorter estimation windows provide better fit in the short term, while larger estimation windows provide robustness in the long run. The optimal balance between these two effects depends on the holding period and in our experiments it is achieved with approximately 3 years observations in the estimation window. Short holding periods do not require robustness in the long run (i.e. shorter estimation windows would deliver better results). As the holding period increases, the long term robustness becomes more relevant than the short term fit and larger estimations windows have to be preferred.

\section{Conclusions}

Portfolio construction is a cornerstone of financial theory and practice. 
However, it is still today a controversial topic for both academics and practitioners. 
Any portfolio optimization strategy relies on assumptions and modeling of the future market structure. 
However, inferring such structure from past observations is a very challenging task, plagued by uncertainty around parameters estimation and relying on some non fully satisfied assumptions. 
\\

We identify three main sources of inaccuracies and errors:
1. model oversimplification;  2. limited size of the estimation set;  3. non-stationarity.
We address oversimplification by introducing a modeling that uses a $L_0$-norm regularized {elliptical} multivariate distribution, demonstrating that it over-performs  traditional models both in likelihood and in portfolio variance performances. We test the effect of sample size by training the models on windows of different sizes and find that performances initially increase with sample size but then eventually decrease for windows above 750 days. We attribute the initial improvement in performance to sampling error, which is reduced when more observations are included, and we interpret the decay in performance when more the 750 observations are included as an instance of non-stationarity. {We further investigate this phenomenon by studying the likelihood corresponding to individual observations out-of-sample and show that shorter estimation windows deliver higher out-of-sample likelihood in the days immediately following the train window, but it tends to rapidly decrease afterwards. As more observations are included in the training set, the out-of-sample likelihood gains stability, with larger values in the long term, but at the cost of lower likelihood in the short term. We conclude that the financial system changes significantly through time and the `optimal' fit in finance needs to be defined in terms of the holding period.}
\\
Our main contribution to the literature on portfolio construction is the demonstration of the relationship between the goodness of the model, measured as out-of-sample likelihood, and the realized portfolio volatility.
We show that higher likelihood obtained with filtered TMFG-LoGo precision matrices correspond to lower portfolio volatility out-of-sample. The relationship between larger likelihood and lower realized volatility is also verified in the maximum-likelihood estimate of the covariance matrix when computed over train sets of different lengths. Further, we show that sparse, filtered covariance matrices can significantly reduce estimation errors coming from both sampling error and non-stationarity.  It also reduces many of the instability problems related to mean-variance optimal weights.

Finally, all of the analysis and conclusions drawn in this paper are based on different estimates of the covariance matrix. While forecasting future returns remains of primary importance in trading and wealth management, we showed that the correlation structure, sometimes overlooked in the asset allocation literature, plays a key role in portfolio construction and a good deal of performances depend upon it.

\section*{Acknowledgments}
TA acknowledges partial support from ESRC (ES/K002309/1),  EPSRC (EP/P031730/1) and EC (H2020-ICT-2018-2 825215).

\bibliographystyle{rQUF}
\bibliography{PortfolioConstruction_SparseMultivariateModelling}

\raggedbottom
\break

\begin{appendices}
\clearpage
\section{Elliptical Distributions}
\label{elliptical_sec}
Consider an $n$-dimensional vector of multivariate returns $\bm{X}=(x_{1},x_{2},...,x_{n})$. If $\bm X$ is elliptical distributed, then its probability density function is defined as:
\begin{equation}
	f_{X}(\bm X) = c_n|\bm{J}|^{1/2} g_n \left[(\bm X-\bm{\mu}) \bm J (\bm X-\bm{\mu})^T\right],
  \label{Elliptical_general}
\end{equation}
where $\bm{\mu}\in \mathbb R^{1\times n}$ is the vector of location (mean) parameters and $c_n$ is a normalization constant. 
The matrix, $\bm J = \bm{\Omega}^{-1} \in \mathbb{R}^{n\times n}$ is the generalized precision matrix, a positively defined matrix which is the inverse of the dispersion matrix $\bm{\Omega}$. When the covariance is defined (as we assume in this paper) then $\bm \Omega = (-\psi'(0))^{-1}\bm \Sigma$, that is, $\bm \Omega$ is proportional to the covariance matrix and the proportionality factor is the inverse of the first derivative of the characteristic generator evaluated at 0.
The function, $g_n(\cdot)$ is called density generator. 

Also, let us stress that $(\bm X-\bm{\mu}) \bm J (\bm X-\bm{\mu})^T$ - i.e. the generalized, square Mahalanobis distance - is a quadratic term and hence a non-negative quantity provided that the matrix $\bm{\Omega}$ is positive definite. To ease the notation, for the remaining of the paper we shall refer to the generalized Mahalanobis distance as $d^2$:
\begin{equation}
	d^2 = (\bm X-\bm{\mu}) \bm J (\bm X-\bm{\mu})^T.
  \label{Mal_distance}
\end{equation}

\par
\bigskip

For dfferent density generators $g_n(\cdot)$ we obtain different distributions of the elliptical family. It is easy to see, for example, that the normal distribution is obtained by using:
\begin{equation}
	g(u) = e^{-u/2},
  \label{normal_generator}
\end{equation}
and $\bm \Omega = \bm \Sigma$.

Similarly the Student-t distribution is obtained by using:
\begin{equation}
	g_{n}(u) = \left(1+\frac{u}{v}\right)^{-\frac{n+v}{2}},
  \label{t_generator}
\end{equation}
where $v$ is the degrees of freedom, and $\bm \Omega = \frac{\nu}{\nu-2}  \bm \Sigma$.

\par
\bigskip
The validity of the \textbf{mean-variance} framework for elliptical distributions has long been established in literature \citep{Owen83}. This proposition is derived easily from two properties of the elliptical distributions. First, for every elliptical distribution with defined mean and variance, the  distribution is completely specified by them (\cite{Owen83} or \cite{Chamberlain83}), with all the higher moments being either zero or proportional to the first or second moment.
Second, any linear combination of multivariate elliptically distributed variables is also an elliptically distributed variable. 
In the case of normal distribution and Stiudent-t distribution they also have the same density generator function. 
Further details on these properties are provided in Appendix~\ref{EllipticalProperties}. 
\par It follows that, if asset returns have a multivariate elliptical distribution $\bm{X} \sim \mathcal{E}_n(\bm{\mu, \Omega}, g_n)$, then the portfolio expected return and dispersion are given by, respectively, $\mathbb{E}[r_p]=\bm{W\mu}^T$ and $\sigma_{p}=\bm{W \Omega W^T}$, matching the optimization framework outlined in Section~\ref{markowitz_sec}.

\par
\bigskip
With respect to our \textbf{likelihood analysis}, considering distributions with probability density function of the form specified in Eq. \ref{Elliptical_general}, the corresponding likelihood function is of the form:
\begin{equation}
	\mathcal{L}_{ED}(\bm{\theta; X}) = \left|\bm{J}\right|^{1/2} \; g_n\left( d^2 \right)\;.
  \label{Elliptical_likelihood}
\end{equation}
Where $ED$ denotes the general Elliptical Distributions and we omitted the constant of integration.\par

To stress the general validity of our analysis for other elliptical distributions, we repeated the experiments discussed in Section~\ref{methodology} considering the t-student generator.

\par
\begin{figure}[H]
\begin{center}
\begin{minipage}{150mm}
\begin{center}
\subfigure[Log likelihood assuming a t-student distribution of log returns.]{
\resizebox*{7cm}{!}{\includegraphics{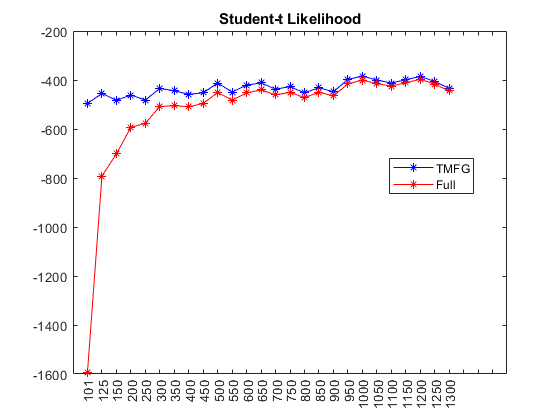}}\label{Student_lik}}
\caption{Out-of-sample log likelihood likelihood computed using the maximum likelihood and the TMFG covariances. These results are coherent with the findings related to the Normal distribution presented in Section~\ref{lik_res}
\label{ED_likelihood}}
\end{center}
\end{minipage}
\end{center}
\end{figure}

\bigskip
Assuming a \textbf{Student - t} distribution of the log returns, the log likelihood (Eq.~\ref{Elliptical_likelihood}) is:
\begin{equation}
	\ln \mathcal{L}_{Student} = \frac{\ln |\bm{J}|}{2} - \frac{n+\nu}{2} \ln \left(1+ \frac{d^2}{\nu-2} \right)
  \label{log_lik_Studentt}
\end{equation}
where $n$ is the sample size and $\nu$ is the degree of freedom. Figure~\ref{Student_lik} reports the likelihood comparison for the same resamplings as in Figure~\ref{Likelihood_comparison} but using a student-t log likelihood as in Eq.~\ref{log_lik_Studentt}. Here we used $n=500$ observations (i.e. the out-of-sample size) and $\nu=3$. We verified  that this findings are robust across different degrees of freedom in the range $\nu=[2.1,4]$.


\section{Properties of Elliptical Distributions}
\label{EllipticalProperties}
In this section we recall some useful properties of Elliptical Distribution which we referred to in our discussion and particularly in Section~\ref{elliptical_sec}.

\begin{property}[\textbf{Distribution Definition}]
Consider an n-dimensional random vector $\bm{X}=(X_1,...,X_n)$. $\bm{X}$ has a multivariate elliptical distribution with location parameter $\bm{\mu}$ and dispersion parameter $\bf \Omega$,written as $\bm{X} \sim \mathcal{E}(\bm{\mu , \Omega} )$ if its characteristic function $\phi$ can be expressed as:
\begin{equation}
	\phi_X(\bm w) = \mathbb{E}(e^{i \bm w \bm{X}})=e^{i \bm w \bm{\mu}}\psi \left({\frac{1}{2}\bm w \bm{\Omega} \bm w^T} \right),
  \label{characteristicF}
\end{equation}
for some location parameter $\bm{\mu} \in \mathbb{R}^{1\times n}  $, positive-definite dispersion matrix $\bm{\Omega} \in \mathbb{R}^{n\times n}$ and for some function $\psi(\cdot):[0,\infty) \rightarrow \mathbb{R}$ such that $\psi \left(\sum_{i=1}^{n}w^2_i \right)$ is a characteristic function, which is called characteristic generator.
If $\bm{X} \sim \mathcal{E}(\bm{\mu , \Omega} )$ and if its density $f_X(\bm{X})$ exists, it is of the form defined in Eq.~\ref{Elliptical_general}.
\end{property}

\bigskip

\begin{property}[\textbf{Density Generator}]
The function $g(\cdot)$ defined in Section~\ref{elliptical_sec} is guaranteed to be density generator if the following condition holds:
\begin{equation}
	\int_{0}^{\infty} x^{n/2-1}g_n(x)dx < \infty.
  \label{characteristicF}
\end{equation}
\end{property}

\bigskip

\begin{property}[\textbf{Affine Equivariance}]

If $\bm{X}=(X_1,...,X_n)$ is an n-dimensional elliptical random variable with location parameter $\bm{\mu}$ and dispersion parameter $\bm{\Omega}$ so that $\bm{X} \sim \mathcal{E}_X(\bm{\mu , \Omega} )$, then for any vector $\bm{a} \in\mathbb{R}^{1\times m} $ and any matrix $\bm{B}\in\mathbb{R}^{m\times n}$ the following affine equivariace holds:
\begin{equation}
	\bm Y = \bm{a + BX}\sim \mathcal{E}_Y(\bm{a+B\mu, B\Omega B}).
  \label{equivariace}
\end{equation}
In other words, any linear combination of multivariate elliptical distributions is another elliptical distribution.
In the special cases of normal, Student-t and Cauchy distributions, the induced density generators are $m$-dimensional version of the original generator of $\bm{X}$. 
\end{property}

For the proof of Properties 1,2 and 3, we refer to \cite{Fang90}.
\par
\bigskip
This implies that any portfolio $Y = \beta_1 X_1+...+\beta_n X_n$ of elliptically distributed variables is distributed accordingly with a (univariate) elliptical distribution, which is a location-scale distribution. 
Furthermore, for any univariate elliptical distribution all moments can be obtained from the first and second moments (if defined). 
In particular, for centered variables with zero mean ($\mu_Y=0$), the resulting distribution of $Y$ is symmetrical around zero and it has all odd moments equal to zero and all even moments given by:
\[
\mu_{2m} = c_m \mu_2^m,
\]
with 
\[
c_m = \frac{(2m)!}{(2^mm!)}\frac{\psi^{(m)}(0)}{(\psi^{(1)}(0))^m}.
\]
Where $\psi^{(m)}(0)$ indicated the $m^{th}$ derivative of $\psi(\omega)$ computed at $\omega=0$.
\par
{As an example, in the normal (0,1) case, $\mu_2=1$, $c_m=0$ for all $m=1,2,...$, the kurtosis is $\mu_{(4)}=\frac{4!}{2^4 4!}=3$, and $\mu_{(2m)}=\frac{(2m)!}{(2^m m!)}$.
For the proof we refer to \cite{Bernake86}, which derived this property by succesive differentiations of $\phi(\cdot)$, and to \cite{Maruyama2003}, which attained the same result by expressing the elliptical distribution in terms of a random vector with uniform distribution on the unit sphere.}
\par 
\bigskip
Therefore the mean-variance optimization is of general applicability and relevance for any portfolio generated from  multivariate elliptically distributed variables.

%
%
%
%
%
%

\end{appendices}

\end{document}